\documentclass[12pt]{article}\pdfoutput=1
\usepackage{cite}
\usepackage{amsmath,amsfonts,amssymb}
\usepackage[small,bf,hang]{caption}
\usepackage{latexsym,epsfig}
\usepackage{arydshln}

\usepackage{tikz-cd}

\usepackage{setspace}

\usepackage{chngcntr}

\def\hybrid{
        \topmargin -20pt
        \oddsidemargin 0pt
        \headheight 0pt \headsep 0pt
       \textwidth 6.5in 
       \textheight 9in 
        \marginparwidth .875in
        \parskip 5pt plus 1pt \jot = 1.5ex}

\hybrid

 \csname
@addtoreset\endcsname{equation}{section}

\def\moth{\mathsurround=0pt}
\newdimen\zo \zo=0pt

\def\tick{\leaders\hrule height 0.5ex depth 0pt \hskip 0.5pt}
\def\upboxfill{$\moth \setbox\zo\hbox{\tick}%
  \hskip 3pt\hbox to 0pt{$\tick$\hss}\hrulefill \hbox to 7.5pt{$\tick$\hss}$}

\def\dtick{\leaders\hrule height .34pt depth 0.5ex \hskip 0.5pt}
\def\downboxfill{$\moth \setbox\zo\hbox{\dtick}%
  \hskip 2pt\hbox to 0pt{$\dtick$\hss}\hrulefill \hbox to 2pt{$\dtick$\hss}$}

\def\cK{{\cal K}}

\def\B{\square}

\def\bpm{\begin{pmatrix}}
\def\epm{\end{pmatrix}}

\def\bec{\begin{center}}
\def\ec{\end{center}}

\def\cN{{\cal N}}

\def\del{\partial}

\def\be{\begin{equation}}
\def\ee{\end{equation}}
\def\bea{\begin{eqnarray}}
\def\eea{\end{eqnarray}}
\def\ba{\begin{array}}
\def\ea{\end{array}}

\begin{document}

\begin{titlepage}
\rightline{}
\rightline{June 2023}
\rightline{HU-EP-23/49-RTG}  
\begin{center}
\vskip 1cm
{\large \bf{Gravity = Yang-Mills}}\\
\vskip 1.2cm

{
\bf {Roberto Bonezzi, Christoph Chiaffrino, Felipe D\'iaz-Jaramillo and Olaf Hohm }}
\vskip 1cm

{\it  Institute for Physics, Humboldt University Berlin,\\
 Zum Gro\ss en Windkanal 6, D-12489 Berlin, Germany}\\[0.5ex]
ohohm@physik.hu-berlin.de, roberto.bonezzi@physik.hu-berlin.de, felipe.diaz-jaramillo@hu-berlin.de, chiaffrc@hu-berlin.de\\ 
\vskip .1cm


\end{center}

\bigskip\bigskip
\begin{center} 
\textbf{Abstract}

\end{center}

\begin{quote}


This essay's title is justified  by 
discussing  a class of Yang-Mills-type  theories of which standard  
Yang-Mills theories are special cases but which is broad enough 
to include gravity as a double field theory. We use the 
framework of homotopy algebras, where conventional Yang-Mills theory 
is the tensor product ${\cal K}\otimes \frak{g}$ of a `kinematic'  algebra ${\cal K}$ with a color Lie algebra $\frak{g}$. 
The larger class of Yang-Mills-type theories are given by 
the tensor product of ${\cal K}$ with more general Lie-type
algebras of which ${\cal K}$ itself is an example, up to anomalies that 
can be cancelled for the tensor product with a second copy $\bar{\cal K}$. 
Gravity is then given by ${\cal K}\otimes \bar{\cal K}$.  


\end{quote}

\vfill

\end{titlepage}



\counterwithout{equation}{section}

\setcounter{equation}{0}

The title of this essay is of course meant as a provocation, a provocation of the kind people engage in when 
they say such things as  ER = EPR ---  statements that  
 appear nonsensical  but that perhaps suggest a 
novel interpretation of the terms involved that could be quite revealing. To say that gravity is a Yang-Mills theory 
is not nonsense  but actually  false if by Yang-Mills theory one means the standard text book theories 
labelled by the structure constants $f_{abc}$ of a `color' Lie algebra $\frak{g}$ and by gravity one means the Einstein-Hilbert 
action in the same dimension. 
Indeed, in the context of perturbative QFT, Yang-Mills theories carry only cubic and quartic couplings and are renormalizable 
while Einstein-Hilbert gravity is non-polynomial and non-renormalizable. 

Rather, here we suggest the definition of a broader class of `Yang-Mills-type' theories, based on the homotopy algebra 
formulation of field theories to be explained below \cite{Zwiebach:1992ie,Lada:1992wc,Hohm:2017pnh}. 
In this framework, textbook Yang-Mills theory takes the form of a tensor product ${\cal K}\otimes \frak{g}$ 
of a homotopy  algebra ${\cal K}$, that encodes the kinematics of Yang-Mills theory, 
with the color  Lie algebra $\frak{g}$ \cite{Zeitlin:2008cc}. 
One may then define the larger  class of  Yang-Mills-type theories as  the tensor product of ${\cal K}$ with more general `Lie-type' algebras, namely homotopy Lie or $L_{\infty}$ algebras. 
The kinematic algebra ${\cal K}$ itself carries, in a hidden way,  such a Lie-type algebra, up to obstructions that, however, 
cancel on a subspace  of the tensor product ${\cal K}\otimes \bar{\cal K}$ with a second copy $\bar{\cal K}$, 
as has been proved  to the order relevant for 
quartic couplings \cite{Bonezzi:2022bse,Bonezzi:2022yuh,Diaz-Jaramillo:2021wtl}. 
We may thus define the yet larger  class of Yang-Mills-type theories as those given by the tensor product of ${\cal K}$
with an obstructed Lie-type algebra for which the obstructions are  cancelled on a  non-trivial subspace. 
So defined, it is just a fact  that there is a  subspace of ${\cal K}\otimes \bar{\cal K}$ defining a Yang-Mills-type theory that 
\textit{is} gravity. Importantly, this theory  is \textit{not} Einstein-Hilbert gravity but rather double field theory which 
 includes the graviton, an antisymmetric tensor (B-field) and a scalar (dilaton) \cite{Hull:2009mi,Siegel:1993th,Hohm:2010pp}.

Our construction is directly inspired by, and a gauge invariant off-shell generalization of, 
the `double copy' technique of  amplitudes \cite{Kawai:1985xq,Bern:2008qj,Bern:2019prr}, 
known to the general public under the slogan `\,Gravity = (Yang-Mills)$^{\,2}$\,'.  
We think 
that the slogan  `\,Gravity = Yang-Mills\,' is more appropriate,  
a viewpoint that can perhaps be justified  most  succinctly as follows: 
Since the standard textbook presentation of Yang-Mills theory gives a Lagrangian in terms of structure constants $f_{abc}$, 
we can  think of Yang-Mills theory as a machine: a machine that takes as input structure constants $f_{abc}$  
and produces as output a QFT. The framework of homotopy algebra allows us to reinvent Yang-Mills theory as a more powerful machine that accepts as input more general `Lie-type' algebras. Feeding-in a conventional Lie algebra this machine produces  standard Yang-Mills theory, but feeding-in a second copy of the kinematic algebra itself produces 
double field theory.

\medskip

Let us begin by explaining the relation between homotopy algebras and field theories. A field theory is defined 
by introducing a set of fields, which here we denote collectively by $\psi$, by specifying field equations or an action and, possibly, by 
specifying gauge symmetries and  their dual Noether/Bianchi identities.  
We will focus on perturbation theory, where the fields form a vector space (the sum of two fields is again a field), 
and so do the gauge parameters, etc. More precisely, the totality of these objects form a \textit{graded} vector space $X$, 
which means that we assign, as a book keeping devise, an integer called degree to each object, depending, e.g.,  on whether it is a field or 
a gauge parameter.  
Such gradings are familiar from the BV-BRST formalism, where they are related to the ghost number \cite{Batalin:1981jr,Henneaux:1989jq}. 

One then defines various maps on this graded vector space in order to encode, for instance,  kinetic terms, interaction vertices, gauge transformations, 
which define the theory. Specifically, at least in perturbation theory one can write the action as 
\begin{equation}\label{firstLinftyAction}
  S  = \frac{1}{2} \big\langle \psi , B_1(\psi)\big\rangle  + \frac{1}{3!} \big\langle \psi, B_2(\psi,\psi)\big\rangle 
  +\frac{1}{4!} \big\langle \psi, B_3(\psi,\psi,\psi)\big \rangle +\cdots\;, 
 \end{equation} 
i.e., as a sum of quadratic terms, cubic terms, and so on. We have written the terms of order $n+1$ in  fields using  a universal 
inner product $\langle\,,\rangle$ and multilinear maps $B_n$ with $n$ arguments. Furthermore, one assumes, for instance,  
that the field equations and gauge transformations with  gauge parameters $\lambda$, respectively, are given by 
 \be\label{EOM}
  B_1(\psi)  +\frac{1}{2} B_2(\psi,\psi)  +\frac{1}{3!} B_{3}(\psi,\psi,\psi)+\cdots \ = \ 0 \;, 
 \ee
and 
 \be\label{deltaPsi} 
  \delta\psi = B_1(\lambda) + B_2(\lambda,\psi) + \frac{1}{2} B_3(\lambda, \psi,\psi)  +\cdots\,. 
 \ee

It must be emphasized that the 
$B_n$,  when evaluated on different objects like fields or gauge parameters, are a priori independent maps, 
distinguished by the degrees of their  inputs. 
For instance, $B_1$, which is known as the differential, acts according to the following chain:  
\begin{equation}\label{bdiagram}
\begin{tikzcd}[row sep=2mm]
\cdots \arrow{r}
&X_{-1}\arrow{r}{B_1} & X_{0}\arrow{r}{B_1} & X_1\arrow{r}{B_1} & X_2  \arrow{r} &\cdots \\
&\lambda& \psi & {\rm EoM}  & {\rm Noether} 
\end{tikzcd}
\end{equation}
where the interpretation of each space is indicated in the second line. 
According to (\ref{deltaPsi}), $B_1$ acting on $\lambda\in X_{-1}$  is defined by the zeroth-order, inhomogeneous terms  
of the gauge transformations, while according to (\ref{EOM}),  $B_1$ acting on $\psi\in X_0$ is defined by the 
linear terms of the equations of motion. For instance, in Yang-Mills theory, $B_1$ on fields is  the second order differential 
`Maxwell operator', while $B_1$ on gauge parameters is the first order differential operator of the abelian gauge transformation.  
Linearized gauge invariance requires 
 \be
  \delta (B_1(\psi)) = B_1(B_1(\lambda)) = 0 \;. 
 \ee 
This relation is summarized as $B_1^2=0$, which moreover must hold on the entire complex (\ref{bdiagram}), 
where it also encodes linearized Noether identities.

More generally, the $B_n$ are determined so that the non-linear equations of motion, gauge transformations and so on 
correctly describe the desired field theory. They must of course satisfy consistency conditions following 
from those of field theory, as for instance gauge covariance of the field equations or closure of the gauge algebra. 
Thus, with any consistent (perturbative) 
field theory we are given   a graded vector space $X$ equipped with 
maps  $B_n$  obeying various consistency relations. This is what mathematicians call a \textit{structure}. 
So what is the general structure of  field theory? This is the category of homotopy Lie algebras 
or  $L_{\infty}$ algebras.   
These generalizations of Lie algebras  are defined as integer graded vector spaces $X=\bigoplus_{i\in \mathbb{Z}}X_i$ equipped with 
multilinear maps $B_n$ of intrinsic degree $+1$ 
(i.e.~the degree of the output is the sum of the input degrees plus 1), which are graded symmetric (e.g., 
$B_2(x_1, x_2) = (-1)^{x_1x_2} B_2(x_2,x_1)$, where in exponents $x$ denotes the degree). 
This means that, depending on the input, the map may be symmetric, as when evaluated on fields, or antisymmetric, 
as when evaluated on gauge parameters. 
The $B_n$ are subject to an infinite number of
 $L_{\infty}$ relations. The first relation is $B_1^2=0$, 
 and we display  the next two relations: 
 \be\label{LinftyREL} 
  \begin{split}
   &B_1(B_2(x_1,x_2)) + B_2(B_1(x_1), x_2) + (-1)^{x_1} B_2(x_1, B_1(x_2))=0  \;, \\[0.8ex] 
   &B_2(B_2(x_1,x_2),x_3) + 
   (-1)^{x_3(x_1+x_2)} B_2(B_2(x_3,x_1),x_2)
   +(-1)^{x_1(x_2+x_3)}  B_2(B_2(x_2,x_3),x_1)  \\
   &+ B_1(B_3(x_1,x_2,x_3)) + B_3(B_1(x_1),x_2,x_3) + \text{two terms} = 0 \;. 
  \end{split}
 \ee   
These state, respectively,  that the differential $B_{1}$ obeys the Leibniz rule with respect to the  2-bracket $B_2$, 
and that the Jacobi identity only needs to hold  `up to homotopy', the failure being governed by the differential $B_1$
and the `3-bracket' $B_3$. The  $L_{\infty}$ relations encode the consistency 
conditions of field theory, for instance the gauge covariance of (\ref{EOM}) under (\ref{deltaPsi}).

We now turn to the $L_{\infty}$ algebra of Yang-Mills theory. The familiar action,  written 
in terms of the non-abelian field strength $F_{\mu\nu}{}^a=\partial_{\mu}A_{\nu}{}^{a}-\partial_{\nu}A_{\mu}{}^{a}+f^{a}{}_{bc} 
A_{\mu}{}^{b} A_{\nu}{}^{c}$, reads 
 \be
  S_{\rm YM}  = -\frac{1}{4}\int dx\,  F^{\mu\nu  a} F_{\mu\nu a} \;, 
 \ee
where $dx$ denotes the flat space volume element in $D$ dimensions. 
Expanded in fields and  using  integrations by part, this becomes 
\begin{equation}\label{YMactionwithphi}
S_{\rm YM} =\int dx \,\Big[\,\tfrac12 A^\mu_a\B A_\mu^a+ \tfrac{1}{2} (\del^\mu A_\mu^a)^2
-f_{abc}\,\del_\mu A_\nu^a A^{\mu b}A^{\nu c}-\tfrac14\,f^e{}_{ab}f_{ecd}\,A_\mu^a A_\nu^b A^{\mu c}A^{\nu d}\, \Big]\;, 
\end{equation}
where $\B=\partial^{\mu}\partial_{\mu}$ denotes the d'Alembert operator. 
For the following applications  it is useful 
to pass to an equivalent formulation by introducing an auxiliary scalar:  
\begin{equation}\label{YMaction}
S_{\rm YM} =\int d x \,\Big[\tfrac12\,A^\mu_a\B A_\mu^a-\tfrac12\,\varphi_a\varphi^a+\varphi_a\, \del^\mu A_\mu^a
+\cdots \Big]\;,  
\end{equation}
with the ellipsis denoting the same cubic and quartic terms as in (\ref{YMactionwithphi}). 
Integrating out $\varphi_a$  one recovers (\ref{YMactionwithphi}).

One could now determine the $L_{\infty}$ algebra of Yang-Mills theory as sketched above, but 
it is more useful to immediately `strip off' color and to write this algebra as a tensor product 
of the color Lie algebra $\frak{g}$ with another kind of homotopy algebra. 
The latter `kinematic' algebra is defined on a vector space ${\cal K}$ that carries the same objects 
as the $L_{\infty}$ algebra of Yang-Mills, but without  color indices, which by an abuse of language 
we denote by the same letters and names. ${\cal K}$ defines a chain complex 
(with degrees shifted by one), 
\begin{equation}\label{NEWbdiagram}
\begin{tikzcd}[row sep=2mm]
&K_{0}\arrow{r}{m_1} & K_{1}\arrow{r}{m_1} & K_2\arrow{r}{m_1} & K_3\\
\cK^{(0)}:&\lambda& A_\mu & E\\
\cK^{(1)}: & &\arrow{ul}{b}\varphi&\arrow{ul}{b}E_\mu &\arrow{ul}{b}\cN
\end{tikzcd} 
\end{equation}
where the differential $m_1$  satisfies  $m_1^2=0$. For instance, on gauge parameters and fields, respectively,  
$m_1$ acts as 
\begin{equation}\label{Qaction}
\begin{split}
m_1(\lambda)&=\bpm\del_\mu\lambda\\\square\lambda\epm\in K_{1}\;, \qquad 
m_1\bpm A_\mu\\\varphi\epm =\bpm\del\cdot A-\varphi\\\square A_\mu-\del_\mu\varphi\epm\in K_2\;,  
\end{split} 
\end{equation}
where we use the short-hand notation 
$\del\cdot A=\partial_{\nu} A^{\nu}$. 
Thanks to the introduction of $\varphi$, 
we have  a $\mathbb{Z}_2$ grading ${\cal K}=\cK^{(0)}\oplus \cK^{(1)}$ and a map $b$ of intrinsic degree $-1$, whose action is indicated 
in the diagram (\ref{NEWbdiagram}).  
For instance, $b(A_{\mu}, \varphi)=\varphi$, where the output is re-interpreted as a gauge parameter and hence degree shifted.   
In addition, ${\cal K}$ carries a graded symmetric 2-product $m_2$ of degree zero and a 3-product $m_3$ of degree $-1$, 
which we display evaluated on fields:  
\begin{equation}\label{m2andm3}
\begin{split}
m^\mu_2(A_1, A_2)&=\del\cdot A_1 A_2^\mu+2\,A_1\cdot\del A_2^\mu+\del^\mu A_1\cdot A_2 - (1\leftrightarrow2)\;,\\
m_3^\mu(A_1,A_2,A_3)&=A_1\cdot A_2\,A_3^\mu+A_3\cdot A_2\,A_1^\mu-2\,A_1\cdot A_3\,A_2^\mu\;,
\end{split}    
\end{equation}
where the external $\mu$ index indicates the vector component in $K_2$. 

A graded vector space ${\cal K}$ equipped with maps $m_1, m_2, m_3$ and possibly higher maps, subject 
to certain symmetry properties,  
is called a 
$C_{\infty}$ algebra (the homotopy version of a \textit{commutative associative} algebra)   
provided that, in addition to $m_1^2=0$, the following relations hold: 
\begin{equation}\label{C-relations}
\begin{split}
m_1(m_2(u_1,u_2))&=m_2(m_1(u_1),u_2)+(-1)^{u_1}m_2(u_1, m_1(u_2) )\;,\\ 
m_2\big(m_2(u_1,u_2), u_3\big)-m_2&\big(u_1,m_2(u_2,u_3)\big)=m_1(m_3(u_1,u_2,u_3))+m_3(m_1(u_1),u_2,u_3)\\
&+(-1)^{u_1}m_3(u_1,m_1(u_2),u_3)+(-1)^{u_1+u_2}m_3(u_1,u_2,m_1(u_3)) \;.
\end{split}    
\end{equation}
The first relation is the Leibniz rule. The second relation states  that $m_2$ is associative `up to homotopy', 
and in general there may be infinitely many more relations. A $C_{\infty}$ algebra is a special case of an 
$A_{\infty}$ algebra where the $m_2$ is graded symmetric while the higher $m_n$ for $n\geq 3$ are  subject 
to graded Young-tableaux-type symmetries.

The $L_{\infty}$ algebra of Yang-Mills theory is now obtained from the $C_{\infty}$ algebra ${\cal K}$ 
by tensoring with the color Lie algebra $\frak{g}$ \cite{Zeitlin:2008cc} (see \cite{Borsten:2021hua} for a review): 
 \be\label{DFTX} 
  X_{\rm YM} = {\cal K}\otimes \frak{g}\;. 
 \ee
At the level of the vector space this just means that the objects in (\ref{NEWbdiagram}) are made $\frak{g}$ valued 
by decorating them with color indices (and degree shifting by one): 
 \be
  x = u^a\otimes t_a \in X_{\rm YM}\;, 
 \ee
where $t_a$ are the generators of  $\frak{g}$. One obtains the fields, gauge parameters, etc., 
of Yang-Mills theory. The 
$B_n$  encoding the $L_{\infty}$ structure on (\ref{DFTX})  are: 
 \be
  \begin{split}
   B_1(x) &= m_1(u^a)\otimes t_a\;, \\
   B_2(x_1,x_2) &= (-1)^{x_1} m_2(u_1^a, u_2^b) f_{ab}{}^{c} \otimes t_c\;, \\
   B_3(x_1,x_2, x_3) &= \Big[(-1)^{x_2} m_3(u_1^a, u_2^b, u_3^c) + (-1)^{x_1(x_2+1)} m_3(u_2^a, u_1^b, u_3^c)\Big] 
   f_{ad}{}^{e} f_{bc}{}^{d} \otimes t_{e} \;, 
     \end{split}
 \ee 
where  in this case 
all $B_n$ for $n>3$ are zero.  
The $L_{\infty}$  relations on $X_{\rm YM}$  follow from the $C_{\infty}$ relations on ${\cal K}$ 
and the Jacobi identities for $f_{ab}{}^{c}$. The resulting $B_2$ and $B_3$ encode the full Yang-Mills theory; 
in particular, using (\ref{m2andm3}), they reproduce the cubic and quartic interactions.

With the decomposition (\ref{DFTX}) of homotopy algebras we have separated Yang-Mills theory into its `kinematic' 
and its `color' parts, with the former being an `associative-type' algebra and the latter a `Lie-type' algebra. 
The idea inspired by the double copy procedure of amplitudes is to replace $\frak{g}$ 
in (\ref{DFTX}) by another type of Lie algebra 
based on the kinematics  of Yang-Mills theory (`kinematic Lie algebra') in order to obtain gravity. 
While ${\cal K}$ started its life as an associative-type algebra, it actually also admits a hidden Lie-type algebra 
(in a suitably generalized sense). 
Borrowing terminology from linguistics we may refer to the former as the `surface structure' and the latter as 
the `deep structure'.

In order to display this deep structure we use  the map $b$ defined in (\ref{NEWbdiagram}), 
which is nilpotent, $b^2=0$, to define a new Lie-type bracket  on ${\cal K}$ 
as the failure of $b$ to act via the Leibniz rule on $m_2$: 
\begin{equation}\label{b2}
b_2(u_1,u_2):=bm_2(u_1,u_2)-m_2(bu_1,u_2)-(-1)^{u_1}m_2(u_1,bu_2)\;.    
\end{equation}
From this definition it follows with $b^2=0$  that $b$ obeys the Leibniz rule 
with $b_2$ and hence $b$ can be thought of as a second differential, of opposite degree to $m_1$.  
The deep structure on ${\cal K}$ is a generalization of a Batalin-Vilkovisky (BV) algebra. 
A BV algebra consists of a (graded) commutative and associative product together with a nilpotent operator 
that, however, does not act via the Leibniz rule on the product but rather  is of `second order' (like the BV Laplacian 
of the BV formalism).  
Defining then a 2-bracket as the failure of the 
differential to act via the Leibniz rule on the product  one obtains a Lie bracket satisfying the Jacobi identities 
and a compatibility condition with the product. The differential $b$, the 2-product $m_2$ and the 
2-bracket $b_2$  above want to be  a BV algebra but fail to be that, because \textit{i)}  $b$ is {not} second order with respect to $m_2$,
and \textit{ii)}  $m_2$ is {not} associative. These failures suggest that there is a  homotopy BV algebra 
(BV$_{\infty}$ algebra \cite{CarrilloVallette}), 
but there is an additional failure due to the relation 
 \be\label{FirstBoxFailure} 
  m_1 b+ b m_1 = \B\,, 
 \ee
where $\B$ denotes the d'Alembert operator.  
This relation quickly follows  with (\ref{NEWbdiagram}), (\ref{Qaction}), and it 
means  that $m_1$ and $b$  are compatible 
only  up to `$\B$--failures'. This has various ramifications. For instance,  
the original differential $m_1$ does not obey the Leibniz rule with respect to $b_2$: 
 \be
  m_1(b_2(u_1,u_2)) + b_2(m_1(u_1), u_2) + (-1)^{u_1} b_2(u_1,m_1(u_2)) =  2\, m_2(\partial^{\mu} u_1, \partial_{\mu} u_2) \,, 
 \ee
where the `anomaly' on the right-hand side follows from (\ref{b2}), (\ref{FirstBoxFailure}) and 
 $\B$ being second order.   
Similar $\B$--failures appear in other relations. Formalizing these failures one can define a more general algebraic structure 
that is realized on the kinematic space ${\cal K}$ of Yang-Mills theory, which following Reiterer 
we denote by  BV$_{\infty}^{\B}$ \cite{Reiterer:2019dys}. 
This  includes as a subalgebra a $C_{\infty}$ algebra and as a `subsector' an $L_{\infty}$ algebra that, however, is 
obstructed by $\B$--failures. [An operator $b$ and an associated  BV$_{\infty}^{\B}$ algebra are also realized 
in self-dual Yang-Mills theory \cite{Bonezzi:2023pox} 
and 3D Chern-Simons theory \cite{Bonezzi:2022bse,Ben-Shahar:2021zww,Borsten:2022vtg}.]

We can now  turn to the construction of gravity in the form of double field theory (DFT). 
Due to the  $\B$--failures, a general BV$_{\infty}^{\B}$ algebra is not quite of `Lie-type' and 
cannot be tensored with the kinematic algebra ${\cal K}$ to obtain a genuine $L_{\infty}$ algebra of gravity. 
However, taking a second copy $\bar{\cal K}$ of the kinematic algebra itself these failures can be cancelled 
on a subspace of the tensor product ${\cal K}\otimes \bar{{\cal K}}$. Denoting all objects of  $\bar{\cal K}$ 
with a bar, the full tensor product space ${\cal K}\otimes \bar{{\cal K}}$ 
is a chain complex carrying two natural differentials of opposite degrees:  
 \be
  B_1:= m_1\otimes {\bf 1} +{\bf 1}\otimes \bar{m}_1\;, \qquad
  b^{-} := \tfrac{1}{2}(b\otimes {\bf 1} -  {\bf 1}\otimes \bar{b})\;, 
 \ee 
which both square to zero due to $m_1^2=b^2=0$. 
The $\B$--failure relation (\ref{FirstBoxFailure})  now implies 
 \be
  B_1 b^- + b^- B_1 = \Delta \;, \qquad \Delta := \tfrac{1}{2} (\B-\bar{\B})\;. 
 \ee 
We can eliminate  the  `$\Delta$--failure'  by going to a subspace with $\Delta=0$. To explain this point we first  note that 
since ${\cal K}$ is a space of functions of coordinates $x$, and $\bar{\cal K}$ is a space of functions 
of  coordinates $\bar{x}$, 
${\cal K}\otimes \bar{\cal K}$ is a space  of functions of doubled coordinates $(x,\bar{x})$. [This is familiar 
from quantum mechanics: the tensor product of two one-particle Hilbert spaces of wave functions of one coordinate 
yields the two-particle Hilbert space of wave functions of two coordinates.] 
We may then impose $\Delta=0$  on functions and products of functions, which in DFT is known as the 
strong constraint and essentially equivalent to identifying 
coordinates $x$ with coordinates $\bar{x}$. 
We will return to the `weakly constrained'  case momentarily.

We thus consider  the subspace 
 \be
 \begin{split}
  &{\cal V}_{\rm DFT} := \Big\{ \psi\in {\cal K}\otimes \bar{\cal K} \,\Big|\, \Delta \psi=0\,,\; b^-\psi=0\Big\}\;, 
  \end{split} 
 \ee
where $b^-\psi=0$ restricts the spectrum appropriately. Together, both constraints in here 
are known as level-matching constraints. The space ${\cal V}_{\rm DFT}$  is precisely 
the complex of DFT as derived from closed string field theory in \cite{Hull:2009mi}. 
For instance, the fields in degree zero (with an overall degree shift by 2) are given by 
 \be
  (e_{\mu\bar{\nu}}\,, \;e\,, \; \bar{e}\,, \; f_{\mu}\,, \; \bar{f}_{\bar{\mu}}) \ \in \ (K_1\otimes \bar{K}_1) 
  \oplus (K_0\otimes \bar{K}_2) \oplus (K_2\otimes \bar{K}_0) \;, 
 \ee
where $e_{\mu\bar{\nu}}$ encodes spin-2 and B-field fluctuations, $e$ and $\bar{e}$ are two `dilatons', one of which 
is pure gauge, and $f_{\mu}$, $\bar{f}_{\bar{\mu}}$ are auxiliary fields that can be integrated out.   
More generally, ${\cal V}_{\rm DFT}$ encodes the gauge parameters  and gauge-for-gauge parameters  of DFT, etc., 
so that, for instance, $\delta \psi = B_1(\Lambda)$ implies 
  $\delta e_{\mu\bar{\nu}} = \partial_{\mu}\bar{\lambda}_{\bar{\nu}}  + \bar{\partial}_{\bar{\nu}}\lambda_{\mu}$,
exhibiting  the `double copy' structure of linearized diffeomorphisms and B-field gauge transformations.

Turning to the non-linear structure we have to define higher brackets on  ${\cal V}_{\rm DFT}$. 
The 2-bracket can be written, in an input-free notation explained in \cite{Bonezzi:2022bse},  as 
\be
B_2:=-\tfrac14\,\big(b_2\otimes\bar m_2-m_2\otimes\bar b_2\big) = -\tfrac{1}{2}\, b^-(m_2\otimes \bar{m}_2)\;,  
\ee
where the second equality holds on the subspace $b^-=0$. 
This 2-bracket  obeys the Leibniz relation, c.f.~(\ref{LinftyREL}), thanks to $b^-$ anticommuting with $B_1$ for $\Delta=0$. 
The second $L_{\infty}$ relation in (\ref{LinftyREL}) involving the Jacobiator  can be satisfied upon defining a suitable $B_3$, 
which is more involved  but can be written entirely in terms of the BV$_{\infty}^{\B}$ 
structures of Yang-Mills theory \cite{Bonezzi:2022bse}. With the above general  $L_{\infty}$  dictionary this determines the 
complete gravity theory to quartic order, in particular, via (\ref{firstLinftyAction}), the quartic couplings. 

Let us emphasize two  features of this construction of gravity  as a Yang-Mills-like theory:
\begin{itemize}
 \item The gauge algebra of DFT, which is  a duality covariant version of the diffeomorphism algebra of gravity, 
 originates rather directly from  the  couplings  of Yang-Mills theory. 
 \item 4-graviton amplitudes can be computed with the $B_2$, $B_3$ above  and by construction  exhibit the 
 factorization into Yang-Mills amplitudes.
\end{itemize}

Above we have essentially  identified the two coordinates $x$ and $\bar{x}$ in order 
to obtain `${\cal N}=0$ supergravity' (Einstein gravity coupled to B-field and dilaton) 
as a  strongly constrained DFT. It is, however, possible to obtain a weakly constrained DFT, 
at least if all dimensions are toroidal, in which the fields genuinely depend on $x$ and $\bar{x}$, subject only 
to $\B=\bar{\B}$ (level-matching for  string theory on tori). 
One uses that the total  space ${\cal K}\otimes \bar{\cal K}$ carries a BV$_{\infty}^{\Delta}$  structure 
and performs an operation known as homotopy transfer (see, e.g., \cite{Vallette,Crainic,Arvanitakis:2020rrk,Arvanitakis:2021ecw}) 
to the subspace $\Delta=0$, together with an additional non-local shift of $B_3$ \cite{Bonezzitoappear}. 
The resulting space realizes an $L_{\infty}$ algebra of weakly constrained functions and hence defines a 
consistent field theory containing  momentum and winding modes without other higher string modes. 
This solves a 
problem that has been open since the seminal work by Hull and Zwiebach \cite{Hull:2009mi}.

We close this essay with a tantalizing possibility. Suppose  a weakly constrained DFT with doubled compact coordinates 
and standard non-compact coordinates can be constructed to all orders in fields. This theory  
is expected to have an  
improved UV behavior.   
While the strongly constrained theory must exhibit the usual UV divergencies of 
general relativity, a weakly constrained theory features infinite towers of additional massive states 
(that in particular are charged under diffeomorphisms along the non-compact dimensions). Since these states 
run in loops, this theory should have an improved UV behavior.  
Indeed, as shown by Sen \cite{Sen:2016qap}, a weakly constrained DFT can in principle 
be derived from the full closed string field theory upon integrating out all string modes that are not part of 
the DFT sector \cite{Arvanitakis:2021ecw}. 
The theory so constructed then inherits the UV finiteness of the full string theory \cite{Sen:2016qap}. 
Therefore, constructing a weakly constrained DFT from scratch might lead, possibly upon including $\alpha'$ corrections \cite{Hohm:2013jaa,Hohm:2014xsa},  to a consistent theory of 
quantum gravity. 
Perhaps there is a quantum theory of gravity much `smaller' than the currently explored string theories, 
and perhaps this quantum gravity is  secretly a Yang-Mills theory. 

\textit{Note added:} After submitting this essay to the arxiv we were informed by Anton Zeitlin 
that his early papers \cite{Zeitlin:2008cc,Zeitlin:2009tj} contain already BV$_{\infty}$ structures 
which, remarkably, have even been suggested to relate to gravity along lines closely related to the
above discussion 
\cite{Zeitlin:2014xma}.

\medskip

 \section*{Acknowledgments} 

We thank Barton Zwiebach for discussions and Anton Zeitlin for correspondence. 

\noindent
This work is funded   by the European Research Council (ERC) under the European Union's Horizon 2020 research and innovation programme (grant agreement No 771862)
and by the Deutsche Forschungsgemeinschaft (DFG, German Research Foundation), ``Rethinking Quantum Field Theory", Projektnummer 417533893/GRK2575.

\end{document}